\documentclass[12pt]{article}

\usepackage{amssymb,amsfonts,amsmath}
\usepackage{graphicx} 
\usepackage{indentfirst}

 \usepackage{bbm}

\parskip=\medskipamount

\newcommand {\cD}{{\cal D}}

\newcommand {\cF}{{\cal F}}
\newcommand {\cG}{{\cal G}}

\newcommand {\cN}{{\cal N}}
\newcommand {\cO}{{\cal O}}

\newcommand {\cV}{{\cal V}}
\newcommand {\cW}{{\cal W}}

\def\a{\alpha}
\def \bi{\bibitem}

\def\b{\beta}

\def\d{\delta}
\def\e{\epsilon}

\def\g{\gamma}
\def\G{\Gamma}

\def\l{\lambda}
\def\m{\mu}

\def\p{\pi}
\def\q{\theta}
\def\r{\rho}

\def\t{\tau}

\def\x{\xi}
\def\z{\zeta}
\def\D{\Delta}
\def\F{\Phi}

\def\Q{\Theta}

\def\U{\Upsilon}

\newcommand{\ve}{\varepsilon}                            

\newcommand{\pa}{\partial}                           
\newcommand{\hf}{\frac12}

\newcommand{\vf}{\varphi}

\newcommand{\be}{\begin{equation}}
\newcommand{\ee}{\end{equation}}
\newcommand{\bea}{\begin{eqnarray}}
\newcommand{\eea}{\end{eqnarray}}
\newcommand{\non}{\nonumber}
\def\double #1{#1{\hbox{\kern-2pt $#1$}}}

\begin{document}

\begin{titlepage}

\begin{flushright}
hep-th/0609078\\
September, 2006\\
\end{flushright}
\vspace{5mm}

\begin{center}
{\large \bf  Five-dimensional
supersymmetric  Chern-Simons action \\
as a hypermultiplet quantum correction }
\end{center}

\begin{center}

{\large  
Sergei M. Kuzenko\footnote{{kuzenko@cyllene.uwa.edu.au}}
} \\
\vspace{5mm}

\footnotesize{
{\it School of Physics M013, The University of Western Australia\\
35 Stirling Highway, Crawley W.A. 6009, Australia}}  
~\\

\vspace{2mm}

\end{center}
\vspace{5mm}

\begin{abstract}
\baselineskip=14pt
\noindent
Building on the  covariant supergraph techniques in 
4D $\cN=2$  harmonic superspace, we develop a manifestly 5D
$\cN=1$ supersymmetric and gauge covariant
formalism to compute the one-loop effective 
action for a hypermultiplet coupled to a background vector multiplet.
As a simple application, we demonstrate the generation of 
a supersymmetric Chern-Simons  action at the quantum level,
both in the Coulomb and the non-Abelian phases.
These superfield results are in agreement with the earlier component
considerations of Seiberg et al.  Our analysis suggests  that similar calculations 
in terms of hybrid 4D superfields or within the 5D projective superspace approach 
may allow one to extract suitable formulations for the non-Abelian
5D supersymmetric Chern-Simons  theory. 
\end{abstract}
\vspace{1cm}

\begin{flushright}
{\it In memory of Steve Irwin}
\end{flushright}

\vfill
\end{titlepage}

\newpage
\setcounter{page}{1}
\renewcommand{\thefootnote}{\arabic{footnote}}
\setcounter{footnote}{0}


${}$Five-dimensional supersymmetric theories with eight supercharges
have recently attracted much attention, mainly in the context of brane-world 
scenarios. Phenomenological applications favour a hybrid superspace formulation 
for these theories in which one keeps manifest only the 4D $\cN=1$ 
supersymmetry, in the spirit of Marcus, Sagnotti and Siegel \cite{MSS}.
A number of 5D rigid supersymmetric models have been constructed 
in such a setting \cite{hybrid1,H,BX}.
Many of them have also been recast in 5D superspace \cite{KL,K1}
(where some new  models have also been put forward).

It seems   robust to view the hybrid and the 5D manifestly supersymmetric 
settings as complimentary to each other.
In the hybrid approach, it is an  open interesting problem
to construct a non-Abelian 5D supersymmetric Chern-Simons 
action.\footnote{Such an action has been given only 
in the Wess-Zumino gauge \cite{Ziegler}.}
In the 5D harmonic superspace approach, such an action 
was constructed several years ago in (the erratum of) \cite{Z}. 
Unfortunately, it is not trivial to reduce the harmonic superspace construction 
of \cite{Z} to the hybrid formalism.  
Still, it is natural to wonder whether 
we can make use of  the construction in \cite{Z} 
to get any practical information, even indirect,
about the explicit structure of the non-Abelian 5D supersymmetric 
Chern-Simon action formulated in terms of 4D $\cN=1$ superfields.
One of the aims of this note is to give a positive answer. 
Ten years ago, it was demonstrated  at the component
level \cite{Witten,Seiberg1,Seiberg2} 
that, in five dimensions, a Chern-Simons term is generated 
by integrating out massive hypermultiplets.
In this note, we follow the ideas put forward in 
\cite{Witten,Seiberg1,Seiberg2}  and  carry out  
explicit one-loop harmonic superspace calculations, 
both along  the Coulomb branch and in the non-Abelian phase, 
and demonstrate that Zupnik's action \cite{Z} appears as 
a leading quantum correction.
Therefore, if one would repeat the same calculations within 
the hybrid superspace formulation,
or within the projective superspace formulation, 
 it should be possible to extract 
a non-Abelian 5D supersymmetric Chern-Simons 
action from the low-energy effective action.

${}$From a more general perspective, this note is aimed at 
developing covariant background field techniques for
computing quantum corrections in 5D $\cN=1$ 
supersymmetric gauge theories. Although there have 
appeared several hybrid superspace calculations
of various quantum effects, see \cite{quantumhybrid,CST} and references therein, 
covariant 5D supergraph techniques seem to be completely 
unexplored, although the properties of  5D $\cN=1$ 
supersymmetric gauge theories are quite interesting
\cite{Witten,Seiberg1,Seiberg2}. Unlike the five-dimensional case, 
powerful covariant supergraph methods have been developed for 
the 4D $\cN=2$ super Yang-Mills theories \cite{BBKO}
(see \cite{BBIKOrev} for a review)  and \cite{KM1}, 
and here we will build  on those results.

The classical action for a massless hypermultiplet 
coupled to a background 5D $\cN=1$ vector multiplet is 
\be
S_{\rm hyper}~ =~
 - \int  {\rm d} \zeta^{(-4)}\,
\breve{q}{}^+ \cD^{++}q^+ ~,
\label{classaction}
\ee
with $\cD^{++} = D^{++} +{\rm i} \, \cV^{++}$
the analyticity-preserving covariant derivative, and 
$\cV^{++}(\z)$ is the analytic prepotential containing all the
information about the off-shell vector multiplet \cite{GIOS}.
The dynamical variable $q^+(\z)$ is a covariantly analytic 
superfield of harmonic U(1) charge +1, $\cD^+_{\hat \a} q^+=0$,
and $\breve{q}{}^+$ is the conjugate of $q^+$ with respect to the 
analyticity-preserving conjugation \cite{GIOS}.
The integration in (\ref{classaction}) is carried out over the analytic subspace 
of the harmonic superspace  ${\mathbb R}^{5|8} \times  S^2$, 
see \cite{KL} for more details 
and our 5D notation and conventions.

The hypermultiplet effective action
reads
\be
\G_{\rm hyper} = {\rm i} \,{\rm Tr}\, \ln \cD^{++}
= - {\rm i} \,{\rm Tr}\, \ln G^{(1,1)}~,
\label{formal}
\ee
with $G^{(1,1)}(\z_1 , \z_2) $  the hypermultiplet Green function
(compare with the four-dimensional case \cite{BBKO}):
\bea
\cD^{++}_1 G^{(1,1)}(\z_1 , \z_2)   &=& \d_A^{(3,1)}(\z_1 , \z_2)
~, \non \\
G^{(1,1)}(\z_1 , \z_2) &=& - \frac{1}{{\stackrel{\frown}{\Box}}{}_1}
(\hat{\cD}_1^+)^4 \,(\hat{\cD}_2^+)^4
\mathbbm{1} 
\delta^{13}(z_1-z_2)
\frac{1}{ (u^+_1 u^+_2)^3}~.
\label{Green}
\eea
Here $\delta^{13}(z-z') =\delta^5(x-x') \, \delta^8(\q-\q')$
is the delta-function in the conventional superspace, 
$ \d_A^{(3,1)}(\z_1 , \z_2)$ the appropriate covariantly analytic 
delta-function, 
\bea
\d_A^{(3,1)}(\z , \z') &=& (\hat{\cD}^+)^4 \,
\mathbbm{1}
\, \delta^{13}(z-z') \,
\d^{(-1,1)}(u,u')~, \non \\
(\hat{\cD}^+)^4 &=& -\frac{1}{ 32}  (\hat{\cD}^+)^2 
\,  (\hat{\cD}^+)^2~, 
\qquad (\hat{\cD}^+)^2 = \cD^{+ \hat \a} \cD^+_{\hat \a}~, 
\eea
and $(u^+_1 u^+_2)^{-3}$ a special harmonic distribution \cite{GIOS}.

In eq. (\ref{Green}),  ${}{\stackrel{\frown} {\Box}}{}$ is the 
covariantly analytic d'Alembertian \cite{KL}
\bea
&&{\stackrel{\frown}{\Box}} = 
\cD^{\hat a} \cD_{\hat a} 
+( {\cal D}^{+ \hat \a}\cW)\,{\cD}^-_{\hat \a}
-\frac{ 1 }{4} 
({\hat \cD}^{+ \hat \a} \cD^+_{\hat \a} \cW )\, \cD^{--}
+\frac{1}{4} ({\cD}^{+ \hat \a}{\cal D}^-_{\hat \a} \cW)
-\cW^2 ~. 
 \eea
Given a covariantly analytic superfield $\vf$, 
$\cD^+_{\hat \a} \vf=0$,
the identity 
$$ {\stackrel{\frown}{\Box}} \,\vf 
=
\hf
({\hat \cD}^+)^4 
(\cD^{--})^2  \vf 
$$ 
holds, and therefore
${\stackrel{\frown}{\Box}} $
preserves analyticity, 
$\cD^+_{\hat \a} \,{\stackrel{\frown}{\Box}} \, \vf=0$.
To prove the above identity, one should make use of 
the following properties
of the 5D $\cN=1$ gauge-covariant derivatives 
in harmonic superspace\footnote{These properties follow from 
the 5D $\cN=1$ vector multiplet formulation \cite{HL}
in  conventional superspace 
 ${\mathbb R}^{5|8}$ parametrized  
by  coordinates  $ z^{\hat A} = (x^{\hat a},  \q^{\hat \a}_i )$.}
\cite{Z,KL}
\bea
\{\cD^+_{\hat \a} \, ,  \cD^-_{\hat \b} \} &=&2 \,
{\rm i}\,  \cD_{\hat \a \hat \b}  
- 2\,\ve_{\hat \a \hat \b} \, \cW  ~,  
\non \\
\big[ \cD^+_{\hat \g} \, ,  \cD_{\hat \a \hat \b} \big] &=&
{\rm i}  \, \Big( \ve_{\hat \a \hat \b} \,\cD^+_{\hat \g } 
+ 2 \ve_{\hat \g \hat \a} \, \cD^+_{\hat \b}
-  2\ve_{\hat \g \hat \b} \, \cD^+_{\hat \a} \Big) \cW~,
\non \\
\big[\cD^{++}\, , \cD^-_{\hat \a} \big] &=&\cD^+_{\hat \a}~, 
\qquad \quad \big[\cD^{++}\, , \cD^+_{\hat \a} \big] =0~,
\non \\
\big[\cD^{--}\, , \cD^+_{\hat \a} \big] &=&\cD^-_{\hat \a}~, 
\qquad \quad \big[\cD^{--}\, , \cD^-_{\hat \a} \big] =0~.
\label{gcd-algebra}
\eea
The field strength $\cW$ obeys the Bianchi identity
\be
\cD^+_{\hat \a} \cD^+_{\hat \b }  \cW
= \frac{1 }{4} \ve_{\hat \a \hat \b} \,
({\hat \cD}^+)^2  \cW \quad 
\Rightarrow \quad 
\cD^+_{\hat \a} \cD_{\hat \b }^+  \cD_{\hat \g }^+ \cW
= 0~.
\label{Bianchi2}
\ee

It should be mentioned that the action (\ref{classaction}) 
is given in the so-called
 $\l$-representation \cite{GIOS} in which the gauge-covariant 
derivatives $\cD^+_{\hat \a}$ possess no connection, i.e. $\cD^+_{\hat \a}$
coincide with the rigid spinor derivatives  
$D^+_{\hat \a}$. 
The explicit expression for the Green function $G^{(1,1)} (\z_1, \z_2)$, 
eq.  (\ref{Green}), is given in the   $\t$-frame \cite{GIOS}
(in the $\l$-frame, the Green function involves the bridge superfield   
at two superspace points \cite{BBKO}).
The $\t$-frame
 is characterised by the properties that (i) the harmonic gauge-covariant 
derivatives $\cD^{++}$ and $\cD^{--}$ possess no connection, 
i.e. $\cD^{\pm \pm }= D^{\pm \pm}$; and (ii) the spinor derivatives 
$\cD^+_{\hat \a}$ and $\cD^-_{\hat \a}$ are expressed in terms of the 
{\it harmonic-independent} gauge-covariant derivatives \cite{HL}, 
\be
\cD_{\hat A} = ( \cD_{\hat a}, \cD_{\hat \a}^i ) 
= D_{\hat A} + {\rm i}\, \cV_{\hat A} (z) ~,
\qquad 
[ \cD_{\hat A} \, ,\, \cD_{\hat B} \} = T_{\hat A \hat B}{}^{\hat C} \, 
\cD_{\hat C}  
+{\rm i} \, \cF_{\hat A \hat B} 
~,
\label{covarder}
\ee 
as follows: $\cD^+_{\hat \a}=\cD^i_{\hat \a} \,u^+_i$
and $\cD^-_{\hat \a}=\cD^i_{\hat \a} \,u^-_i$.
Here $D_{\hat A} = ( \pa_{\hat a}, D_{\hat \a}^i ) $
are the flat covariant derivatives obeying 
the anti-commutation relations
$[ D_{\hat A} , D_{\hat B} \} = T_{\hat A \hat B}{}^{\hat C} \, 
D_{\hat C} $. 

The above definition of $\G_{\rm hyper} $
can be seen  to be purely formal, 
since the operator $\cD^{++}$ maps analytic superfields
$q^+$ with U(1) charge +1 to analytic superfields of 
U(1) charge +3, see  also \cite{KM1}.
However, the expression for an arbitrary variation
of the effective action
\be
\d \G_{\rm hyper} = - {\rm Tr}\,\Big\{\d \cV^{++} \,G^{(1,1)}\Big\}
\label{var1}
\ee
can be made well-defined.
Using Schwinger's proper-time
representation \cite{Schwinger},
we introduce a regularized variation of the effective action
\bea
\d\G_{{\rm hyper}, \e} &=&{\rm tr} 
\int {\rm d} \z^{(-4)}\, \d \cV^{++} 
\langle J^{++}_\e \rangle_\l ~,  \qquad \qquad \quad
\cD^+_{\hat \a} \langle J^{++}_\e \rangle=0~,
\non \\
\langle J^{++}_\e \rangle_\t & =&
\int_{0}^{\infty}{\rm d}({\rm i}s)\,
({\rm i}\,\m^2 s)^\e  \,
{\rm e}^{{\rm i}\,s\,{\stackrel{\frown}{\Box}}_1} \,
(\hat{\cD}_1^+)^4 \,(\hat{\cD}_2^+)^4 \,
\frac{ \mathbbm{1}
 \,\delta^{13}(z_1-z_2)}{ (u^+_1 u^+_2)^3 }
\Big|_{1 = 2}~,
\label{current1}
\eea
with $\e$ the regularization parameter,
$\e \to 0$ upon renormalization,
and $\m$ the normalization point.
Gauge invariance of the effective action is equivalent 
to the fact that $ \langle J^{++}_\e \rangle$ is a conserved current, 
\be
\cD^{++} \langle J^{++}_\e \rangle =0~.
\ee
The second line in  (\ref{current1})
can be brought to a more
useful form by applying the identity
\bea
 (\hat{\cD}^+_1)^4 (\hat{\cD}^+_2)^4 \,
\frac{
\mathbbm{1}
\d^{13} (z_1 -z_2)  }{(u^+_1 u^+_2)^q} &=&
(\hat{\cD}^+_1)^4 \;
\Bigg\{ (\hat{\cD}^-_1)^4 \;
\frac{1 }{(u^+_1 u^+_2)^{q-4}}
- \frac{1}{4} \,
\D^{--}_1\;
\frac{(u^-_1 u^+_2) }{(u^+_1 u^+_2)^{q-3}} \non \\
- {\stackrel{\frown}{\Box}}_1 \;
\frac{(u^-_1 u^+_2)^2 }{(u^+_1 u^+_2)^{q-2}}
&-& \frac{1}{4}(q-3)\;  (\cD^+_1 \cD^+_1 \cW_1)\;
\frac{(u^-_1 u^+_2)^3 }{(u^+_1 u^+_2)^{q-1}} \Bigg\}\,
\mathbbm{1}
\,\d^{13} (z_1 -z_2) ~,
\label{master}
\eea
with $q$ an integer,
and therefore
\bea
 (\hat{\cD}^+_1)^4 (\hat{\cD}^+_2)^4 \,
\frac{
\mathbbm{1}
\d^{13} (z_1 -z_2)  }{(u^+_1 u^+_2)^3} &=&
(\hat{\cD}^+_1)^4 \,
\Bigg\{ (\hat{\cD}^-_1)^4 \,
(u^+_1 u^+_2)
- \frac{1}{4} \,
\D^{--}_1\;
{(u^-_1 u^+_2) }
\non \\
&&\qquad \qquad - {\stackrel{\frown}{\Box}}_1 \;
\frac{(u^-_1 u^+_2)^2 }{(u^+_1 u^+_2)}
\Bigg\}\,
\mathbbm{1}
 \,\d^{13} (z_1 -z_2) ~.
\label{master3}
\eea
Here
\bea
\D^{--} ={\rm i}\,\cD^{\hat \a \hat \b} \cD^-_{\hat \a}  \cD^-_{\hat \b}
&+&  \cW (\hat{\cD}^-)^2 
+ 4 (\cD^{- \hat \a} \cW) \cD^-_{\hat \a}  
+ (\cD^- \cD^- \cW)~.
\eea
The identity (\ref{master}) is a five-dimensional analogue 
of the one obtained in \cite{KM1}. 
It can be derived by using
the anti-commutation relations (\ref{gcd-algebra}). 

Similar to the four-dimensional case \cite{BBKO}, 
the operator
${\stackrel{\frown}{\Box}}$ 
possesses the property
\bea 
(\hat{\cD}^+)^4 \, {\stackrel{\frown}{\Box}}
&=& {\stackrel{\frown}{\Box}} \, (\hat{\cD}^+)^4~. 
\label{prop1}
\eea
Also in complete analogy with  the four-dimensional case \cite{KM1},
one can show that the third term in (\ref{master3})
does not contribute to 
$\langle J^{++}_\e \rangle$ in the limit $\e \to 0$.
Therefore, the current in (\ref{current1}) 
can be rewritten as 
follows\footnote{By construction, the first line of eq. (\ref{current1}) is given in the 
 $\l$-representation.
When computing $\langle J^{++}_\e \rangle$ 
using eq. (\ref{current2}), we 
switch from the $\l$-frame to the $\t$-frame.}
\bea
\langle J^{++}_\e \rangle_\t & =&
\int_{0}^{\infty}{\rm d}({\rm i}s)\,
({\rm i}\,\m^2 s)^\e  \,
(\hat{\cD}^+)^4 \,
{\rm e}^{{\rm i}\,s\,{\stackrel{\frown}{\Box}}} \non \\
&& \qquad \quad \times
\Big\{ (u^+ u'^{+})\,
(\hat{\cD}^-)^4 
+ \frac{1}{4} \,
\D^{--}
\Big\} 
\mathbbm{1}
\,\delta^{13}(z-z')
\Big|_{z = z', \,u=u'}~,
\label{current2}
\eea
where we have use the indentity $(u^+u^-)=-(u^-u^+)=1$.

At this stage, it is useful to introduce, following the general approach
developed in \cite{KM2}, 
a new representation for the full delta-function
\bea
\mathbbm{1}
\,\delta^{13}(z-z') &=& I(z,z') \,\delta^5(x-x') \, \delta^8(\q-\q')
=I(z,z')  \int \frac{ {\rm d}^5 k }{(2\p)^5}\,
{\rm e}^{{\rm i} \,k_{\hat a} \r^{\hat a}} \, \delta^8(\Q) ~,
\eea
where
\bea
 \x^{\hat A}  \equiv \x^{\hat A} (z,z') = -\x^{\hat A} (z',z) 
= \left\{
\begin{array}{l}
\r^{\hat a} = (x-x')^{\hat a} 
+ {\rm i}\, \q^{\hat \a}_i (\G^{\hat a})_{\hat \a \hat \b}\, \q'^{\hat \b i}
 ~, \\
\Q^{\hat \a}_i = (\q - \q')^{\hat \a}_i 
\end{array} 
\right. 
\label{super-interval}
\eea
is the supersymmertic two-point function, and $I(z,z')$ stands for the 
parallel displacement propagator along the straight line connecting 
the points $z$ and $z'$.
The latter is a unique two-point function, which  takes 
its values in the gauge group and 
which obeys the first-order differential equation equation  
and special boundary condition
\be
\x^{\hat A} \cD_{\hat A} \, I(z,z') 
= \x^{\hat A} \Big( D_{\hat A} +{\rm i} \, \cV_{\hat A}(z) \Big) I(z,z') =0~, 
\qquad I(z,z) = \mathbbm{1}
~.
\label{super-PDO2}
\ee
These imply the important relation
\be
I(z,z') \, I(z', z) = \mathbbm{1}
~,
\ee
and also the equation at $z'$
\be
\x^{\hat A} \cD'_{\hat A} \, I(z,z') 
= \x^{\hat A}  \Big( D'_{\hat A} \,I(z,z') 
 - {\rm i} \, I(z,z') \, \cV_{\hat A}(z') \Big) =0~.
\ee
One of the fundamental properties of $I(z,z')$  \cite{KM2} is 
\bea 
\cD_{\hat B} I(z,z') &=& {\rm i} \, I(z,z') \,
\sum_{n=1}^{\infty} \frac{ 1  }{ (n+1)!} \,
\Big\{
n \, \x^{{\hat A}_n} 
\ldots \x^{{\hat A}_1}  
\cD'_{{\hat A}_1} \ldots \cD'_{{\hat A}_{n-1} } \cF_{{\hat A}_n \,{\hat B} } (z') 
\label{super-PTO-der1} \\
&+& \hf  
(n-1)\, 
\x^{{\hat A}_n} T_{{\hat A}_n \,{\hat B}}{}^{\hat C} \,\x^{{\hat A}_{n-1}} 
\ldots \x^{{\hat A}_1}  
\cD'_{{\hat A}_1} \ldots \cD'_{{\hat A}_{n-2} } \cF_{{\hat A}_{n-1} \,{\hat C} } (z') \Big\}~.
\non 
\eea
Together with the identity
\be
\cD^i_{\hat \a} \,\r^{\hat a} = - {\rm i}\,  (\G^{\hat a})_{\hat \a \hat \b}\, \Q^{\hat \b i}
\quad \longrightarrow \quad 
\x^{\hat B}\, \cD_{\hat B} \x^{\hat A} = \x^{\hat A}~,
\ee
eq. (\ref{super-PTO-der1})
 allows us to compute the integrand in (\ref{current2})
in a manifestly covariant way, as a series in powers of the field strength 
and its covariant derivatives.\footnote{Somewhat different 
techniques, valid specifically for one-loop calculations, 
 were suggested  in \cite{Gauss}.}
As a first step, one pushes
the plane wave $\exp ({\rm i} \,k_{\hat a} \r^{\hat a})$ through
all the operatorial factors in (\ref{current2}) to the left, and then 
it turns into unity in the coincidence limit. This has the following impact 
on the covariant derivatives
\be
\cD_{\hat a} ~\to ~\cD_{\hat a}+ {\rm i} \,k_{\hat a} ~,
\qquad 
\cD_{\hat \a}^i ~\to ~\cD_{\hat \a}^i
+ k_{\hat a} \, (\G^{\hat a})_{\hat \a \hat \b}\, \Q^{\hat \b i}~.
\ee
Then, the momentum integration reduces to doing Gaussian integrals
\be
\frac{1}{ (2 \p )^5} \,\int {\rm d}^5 k \,
{\rm e}^{- {\rm i} \,s \,k^2  
} \,
s k^{{\hat a}_1} \ldots s k^{\hat{a}_n}
=\frac{1}{ (4 \p s )^{5/2}} \,\int {\rm d}^5 k \,
{\rm e}^{- {\rm i} \,k^2  
} \,
\sqrt{s} k^{{\hat a}_1} \ldots \sqrt{s} k^{\hat{a}_n}
~,
\label{moment}
\ee
and this is a textbook problem. Finally, the covariant derivatives in 
(\ref{current2}) should hit either $\x^{\hat A} $ (this is easy) or 
the parallel displacement propagator, and then eq.  (\ref{super-PTO-der1})
applies.

We are now prepared to compute the effective action.
Consider first the Coulomb branch of the theory,
where the gauge field takes its values in the Cartan subalgebra.
Let us denote ${\stackrel{\frown}{\Box}} =\cO -\cW^2$. Then
\bea
{\rm e}^{{\rm i}\,s\,{\stackrel{\frown}{\Box}}} 
&=&{\rm e}^{-{\rm i}\,s\,\cW^2}\,{\rm e}^{{\rm i}\,s\,\cO} 
~+~O(\cD\,\cW^2) \\
&=& {\rm e}^{-{\rm i}\,s\,\cW^2}
{\rm e}^{{\rm i}\,s \cD^{\hat a} \cD_{\hat a}} 
\Big\{ 
1 -\frac{ {\rm i}\,s }{4} 
(\cD^{+ } \cD^+ \cW )\, \cD^{--}
+\frac{({\rm i}\,s)^2}{2} \,( {\cal D}^{+ \hat \b}\cW)
( {\cal D}^{+ \hat \a}\cW)\,{\cD}^-_{\hat \a}{\cD}^-_{\hat \b} \Big\} +\dots
\non
\eea
The second term in the curly brackets produces a non-vanishing result 
when hitting $(u^+u'^+) $ in (\ref{current2}), 
$$
\cD^{--} (u^+u'^+)|_{u=u'} = (u^-u'^+)|_{u=u'} =-1~.
$$
The third term in the curly brackets can produce a non-vanishing 
contribution to (\ref{current2}) only if paired with  $\cW (\hat{\cD}^-)^2 $ in 
$\D^{--}$, 
due to the dentity ${\cD}^-_{\hat \a}{\cD}^-_{\hat \b} (\hat{\cD}^-)^2 
=-8\,\ve_{\hat \a \hat \b} \,(\hat{\cD}^-)^4.$

The result of calculation is 
\bea
 \langle J^{++}_\e \rangle &=& {\rm i}
\int_{0}^{\infty}
{\rm d}s \, \frac{(\m^2 s)^{\e} }{(4\p s)^{5/2}}\,
\Big\{
\frac{1 }{4} 
(\cD^{+ } \cD^+ \cW )\, s
-\cW\,
( {\cal D}^{+ \hat \a}\cW)
( {\cal D}^+_{ \hat \a}\cW)
\, s^2 \Big\} \,{\rm e}^{-s \cW^2  }+\dots
\eea
Using the identities $\G(1/2)= (-1/2)\,\G(-1/2) =\sqrt{\p}$, we obtain in the limit 
$\e \to 0$
\bea
 \langle J^{++} \rangle =- {\rm sign} (\cW)\,
 \frac{ 1 }{2(4\p )^2} \, \cG^{++}+\dots~,
\label{abelaincurrent} 
\eea
where $\cG^{++}$
denotes the covariantly 
analytic descendant of $\cW$ introduced in \cite{KL}
\be
-{\rm i} \, \cG^{++} = 
\cD^{+ \hat \a} \cW \, \cD^+_{\hat \a} \cW 
+\frac{1 }{4} \{\cW \,, 
({\hat \cD}^+)^2 \cW \}~, \qquad 
\cD^+_{\hat \a} \cG^{++} = 
 \cD^{++} \cG^{++} =0~.
\label{YML}
\ee
This expression for $\cG^{++}$ holds for the general  non-Abelian 
vector multilpet. When the vector multiplet is restricted to the
Cartan subalgebra, i.e. the case we are currently discussing, 
the anticommutator $\{\cW \,, ({\hat \cD}^+)^2 \cW \}$
in (\ref{YML})
reduces to $2\,\cW  ({\hat \cD}^+)^2 \cW $.

In the U(1) case, the supersymmetric  Chern-Simons action, 
which was first constructed in \cite{Z} in terms of the prepotential 
$\cV^{++}$, can be represented in the form 
\cite{KL}
\be
S_{\rm CS} =  \frac{1}{3 }
\int {\rm d} \z^{(-4)} \, \cV^{++} \,
\cG^{++} ~.
\label{CS2}
\ee
Varying $S_{\rm CS}$ gives
\be
\d S_{\rm CS} =  
\int {\rm d} \z^{(-4)} \, \d \cV^{++} \,
\cG^{++} ~.
\label{CS-var}
\ee
Comparing with (\ref{abelaincurrent}) we see that the leading quantum 
correction computed indeed coincides with a sum of  
super Chern-Simons actions associated with all the  U(1) factors  
in the Coulomb branch.

In the non-Abelian case, the supersymmetric  Chern-Simons action will be 
defined  to vary as follows:
\be
\d S_{\rm CS} =  
{\rm tr}
\int {\rm d} \z^{(-4)} \, \d \cV^{++} \,
\cG^{++} ~,
\label{CS-var2}
\ee
similarly to (\ref{CS-var}).
This definition is  equivalent to the one originally  given in \cite{Z}. 
Indeed, in the $\l$-frame the field strength $\cW$ has a simple expression 
in terms of $\cV^{--}$
\be
\cW_\l= \frac{\rm i}{8} (\hat{D}^+)^2 \cV^{--}~,
\ee
and this can be used to  show that 
\be
\cG^{++}  = \hf (\hat{D}^+)^4 \{ \cW_\l, \cV^{--} \}~.
\ee
Then, eq. (\ref{CS-var2}) can be rewritten
\be
\d S_{\rm CS} =  \hf \,
{\rm tr}
\int {\rm d}^{13}z \,{\rm d}u \, \d \cV^{++} \,
 \{ \cW_\l, \cV^{--} \}~,
\label{CS-var3}
 \ee
 what coincides with the definition given in \cite{Z}.
Zupnik has integrated the variation (\ref{CS-var3}) 
and derived $S_{\rm CS}$ as an infinite series in powers of 
the prepotential $\cV^{++}$. This series terminates 
if one chooses a standard Wess-Zumino gauge for the vector multiplet, 
and then the action can be readily reduced to components.

Instead of giving the explicit expression for $S_{\rm CS}$
in terms of the prepotential,
let us simply demonstrate the integrability of (\ref{CS-var3}).
 Consider a second variation
\be
\d_2 \d_1 S_{\rm CS} =  \
{\rm tr}
\int {\rm d}^{13}z \,{\rm d}u \, \d_1 \cV^{++} \,
 \{ \d_2 \cV^{--} ,\cW_\l \}~,
\label{CS-var4}
 \ee
 and transform it into the $\t$-frame
 \bea
\d_2 \d_1 S_{\rm CS} &=&  
{\rm tr}
\int {\rm d}^{13}z \,{\rm d}u \, (\d_1 \cV^{++})_\t \,
 \{( \d_2 \cV^{--})_\t ,\cW \} \non \\
& =&  
{\rm tr}
\int {\rm d}^{13}z \,{\rm d}u \, \cW\, \{ (\d_1 \cV^{++})_\t \,
 , ( \d_2 \cV^{--})_\t \} ~.
\label{CS-var5}
 \eea
In the $\l$-frame, the  variations $\d V^{++} $ and $\d V^{--}$ 
are related to each other as follows \cite{Z2}
\be
\cD^{++} \d \cV^{--} = \cD^{--} \d \cV^{++}~.
\ee
In the $\t$-frame, this becomes 
\be
D^{++} (\d \cV^{--})_\t = D^{--} (\d \cV^{++})_\t~.
\ee
This equation is known to have the following  solution 
\cite{GIOS,Z2}
\bea
(\d \cV^{--})_\t (u) &=& 
\int  {\rm d} u_1 \,
\frac{
(\d \cV^{++})_\t (u_1)}
{ (u^+ u^+_1)^2 }~.
\eea
Using this result in (\ref{CS-var5}) and taking into account the fact that 
$\cW$ is harmonic-independent, in the $\t$-frame, we conclude
\be
\d_2 \d_1 S_{\rm CS} =\d_1 \d_2 S_{\rm CS} ~.
\ee

Now, let us turn to the consideration of a massive hypermultiplet
transforming in an arbitrary representation of the gauge group $G$.
One can use the same action (\ref{classaction}) provided 
one assumes that (i) the gauge group is $G \times {\rm U(1)}$, 
and (ii) the  U(1) gauge field $\cV^{++}_0$ possesses 
a constant field strength $\cW_0= {\rm const}$,  
$|\cW_0|=m$, see \cite{vev} for more details.
This effectively amounts to replacing 
\be 
\cV^{++} ~\to ~ \tilde{\cV}^{++} =\cV^{++}_0 +\cV^{++}~, 
\qquad \cW ~\to ~ \tilde{\cW} =\cW_0 +\cW
\ee
in  most of the above expressions. Of course, we should also 
modify the gauge covariant derivatives and 
the field strength $\cF_{\hat A \hat B} $ 
in (\ref{covarder}) similarly. 
The  U(1) gauge field $\cV^{++}_0$
is completely frozen, and therefore eq. (\ref{var1}) involves only 
the variation of $\cV^{++}$ corresponding 
to the actual gauge group $G$.
With all such modifications in mind, eq. (\ref{current2})
still holds, and we can use it for computing the variation 
of the effective action. 

Since we now have a large mass parameter 
in the theory, the effective action can be computed 
in the most traditional manner, as an expansion in inverse powers of $m$, 
with the generic non-Abelian gauge field.
Let us represent 
${\stackrel{\frown}{\Box}} =\tilde{\cO} -(\cW_0 +\cW)^2
=\cO - \cW_0{}^2= \cO - m^2$.
Then we can represent the operator 
${\rm exp}({\rm i}s\,{\stackrel{\frown}{\Box}}) $ in  (\ref{current2}) as
\bea
{\rm e}^{{\rm i}\,s\,{\stackrel{\frown}{\Box}}} 
&=& {\rm e}^{-{\rm i}\,s\,m^2}
{\rm e}^{{\rm i}\,s \cD^{\hat a} \cD_{\hat a}} 
\Big\{ 
1 -\frac{ {\rm i}\,s }{4} 
(\cD^{+ } \cD^+ \cW )\, \cD^{--}
+\frac{({\rm i}\,s)^2}{2} \,( {\cal D}^{+ \hat \b}\cW)
( {\cal D}^{+ \hat \a}\cW)\,{\cD}^-_{\hat \a}{\cD}^-_{\hat \b} \Big\} +\dots
\non
\eea
This will lead to 
\bea
 \langle J^{++}_\e \rangle &=& {\rm i}
\int_{0}^{\infty}
{\rm d}s \, \frac{(\m^2 s)^{\e} }{(4\p s)^{5/2}}\,
\Big\{
\frac{1 }{4} 
(\cD^{+ } \cD^+ \cW )\, s
- ( {\cal D}^{+ \hat \a}\cW)
( {\cal D}^+_{ \hat \a}\cW)\,(\cW_0 +\cW)
\, s^2 \Big\} \,{\rm e}^{-s \,m^2  }+\dots \non \\
&=& - {\rm sign} (\cW_0)\,
 \frac{ \rm i }{2(4\p )^2} \, 
\Big\{ \hf  (\cD^{+ } \cD^+ \cW )\, \cW_0
+( {\cal D}^{+ \hat \a}\cW)
( {\cal D}^+_{ \hat \a}\cW)\,(\cW_0 +\cW) \,\cW_0{}^{-1} \Big\}
+\dots
\non
\eea
By construction, the effective action should depend on $\cW_0$ 
only through the combination $(\cW_0 +\cW)$. However, this structure 
has been spoiled by our calculational scheme 
which requires us to Taylor expand contributions with $(\cW_0 +\cW)^2$
at the point $\cW_0^2 =m^2$.
But the same scheme 
clearly indicates how one can  restore the required structure
in the final expression.
This is similar to the approach used in 
\cite{BO}.  We end up with 
\bea
 \langle J^{++}_\e \rangle &=&
- |\cW_0| \, \frac{ \rm i }{64\p^2} \,   (\cD^{+ } \cD^+ \cW )
- {\rm sign} (\cW_0)\,
 \frac{ 1 }{32\p^2} \, \cG^{++} +\dots~,
\label{non-abelaincurrent}
\eea
with $\cG^{++}$ defined in (\ref{YML}).
As in \cite{BO}, our final result 
lacks uniqueness to the extent that we ignore some commutator 
terms which should be treated as higher order quantum corrections.

The first term on the right of eq. (\ref{non-abelaincurrent}) 
generates the super Yang-Mills term, see \cite{KL} for the relevant details, 
while the second term corresponds to  the super Chern-Simons action, 
in accordance with our previous discussion.

The one-loop calculation performed can also be  
carried out  using either the hybrid superspace approach 
or   projective supergraph techniques \cite{G-R}.
The hypermultiplet action in 4D superspace
has the Fayet-Sohnius form \cite{hybrid1}
\bea
S&=&
\int {\rm d}^5 x \left\{ \int {\rm d}^4 \q \,( 
Q^\dagger   {\rm e}^{\cV}  Q 
+ \tilde{Q}  \, {\rm e}^{-\cV}  \tilde{Q}^\dagger )
+ \Big( \int {\rm d}^2 \q\,  	
\tilde{Q} (\F +m  - \pa_5 ) Q
+{\rm c.c.}\Big) \right\} ~.
\label{FS-actionN=1}
\eea
Here the chiral superfields $Q$ and $\tilde{Q}$ describe the 
hypermultiplet, while the adjoint gauge $V$ and chiral $\F$
superfields correspond to the background 5D vector multiplet.
To compute the hypermultiplet effective action, one 
can use powerful 4D $\cN=1$ functional techniques, 
see e.g. \cite{BK}.
The hypermultiplet action in 5D projective superspace \cite{KL} is
\be
S= \frac{1}{2\pi {\rm i}} \, \oint \frac{{\rm d}w}{w} 
 \int {\rm d}^5 x \, {\rm d}^4 \q \,  
 \breve{\U} (w)\, {\rm e}^{\cV(w) +\cV_0(w)} \, \U(w)  \ ~.
\label{projhyper}
\ee
Here the hypermultiplet is described by an arctic superfield
$\U(w) $ and its conjugate, 
and the 5D vector multiplet is described by a tropical superfield
$\cV(w)$, see \cite{G-R,KL} for more details.

Recently, it has been claimed \cite{GPT} that  projective supergraph 
techniques are more efficient than the harmonic ones.
So far, this claim does not seem to have much evidence to support it.
Conceptually, the projective supergraphs are essentially equivalent
to the harmonic ones
\cite{K-pro}. 
In terms of factual evidence, 
a great many  covariant supergraph calculations 
for 4D $\cN=2$ super Yang-Mills theories 
have been carried out within the harmonic superspace approach, 
see e.g. \cite{BBKO,BBIKOrev,KM1}, whereas there has appeared 
 only one non-trivial calculation \cite{G-RR} based on the 
use of   the projective supergraphs. Therefore, it would be very interesting 
to compute a one-loop low-energy effective action for the theory
(\ref{projhyper}) and try to extract from it a non-Abelian supersymmetric 
Chern-Simons action realized in terms of 
the tropical prepotential $\cV(w)$ (the Abelian version was given in \cite{KL}).\\

\noindent
{\bf Acknowledgements:}\\
It is a pleasure to thank Jim Gates, Arthur Hebecker, Shane McCarthy 
and Robert Ziegler for conversations about 5D super Chern-Simons  
actions, and Ian McArthur for reading the manuscript. This work  is supported  
by the Australian Research Council and by a UWA research grant.

\small{

}

\end{document}